\documentclass[journal]{IEEEtran}

\usepackage{booktabs}
\usepackage{multirow}
\usepackage{algorithm,algorithmic}
\usepackage{amsmath}
\usepackage{mathtools}
\renewcommand{\baselinestretch}{0.95} 

\usepackage{multicol}

\usepackage{ifpdf}

\usepackage{cite}

\ifCLASSINFOpdf
  \usepackage[pdftex]{graphicx}
\else
\fi
\usepackage{amsmath}
\usepackage{algorithmic}

\usepackage{array}

\ifCLASSOPTIONcompsoc
  \usepackage[caption=false,font=normalsize,labelfont=sf,textfont=sf]{subfig}
\else
  \usepackage[caption=false,font=footnotesize]{subfig}
\fi
\usepackage{fixltx2e}

\usepackage{url}

\usepackage{dblfloatfix}    

\usepackage{kantlipsum}     

\begin{document}
%
\title{Security Protection for Magnetic Tunnel Junction}


\author{
\IEEEauthorblockN{Shayan Taheri and Jiann-Shiun Yuan}\\
\IEEEauthorblockA{Department of Electrical and Computer Engineering, University of Central Florida, Orlando, FL 32816, U.S.A.\\
Email: shayan.taheri@knights.ucf.edu, jiann-shiun.yuan@ucf.edu}
}


%


\maketitle

\begin{abstract}

Energy efficiency is one of the most important parameters for designing and building a computing system nowadays. Introduction of new transistor and memory technologies to the integrated circuits design have brought hope for low energy very large scale integration (VLSI) circuit design. This excellency is pleasant if the computing system is secure and the energy is not wasted through execution of malicious actions. In fact, it is required to make sure that the utilized transistor and memory devices function correctly and no error occurs in the system operation. In this regard, we propose a built-in-self-test architecture for security checking of the magnetic tunnel junction (MTJ) device under malicious process variations attack. Also, a general identification technique is presented to investigate the behavior and activities of the employed circuitries within this MTJ testing architecture. The presented identification technique tries to find any abnormal behavior using the circuit current signal.

\end{abstract}






%
\IEEEpeerreviewmaketitle

\section{Introduction}



The ubiquitous connectivity among computing systems is increasing and consequently significant growth is happening in the amount of data to be processed, transmitted, and stored by these systems. This situation brings a proper environment for adversaries to exploit possible backdoors in software and/or hardware to perform malicious purposes. Besides security, another design parameter that is highly critical for computing systems, especially in mobile devices, is energy. The dream of building a smart city with having millions of electronic devices around us is not possible, unless making them energy efficient.

Recently, new transistor and memory technologies are introduced to the very large scale integration (VLSI) circuit design for the sake of low energy consumption, especially due to the device scaling barriers of the CMOS technology. These devices such as magnetic tunnel junction (MTJ) and tunnel field-effect transistor (TFET) are able to drop the energy consumption of electronic circuits remarkably. Although their merit is not only limited to energy reduction since they have unique features and properties applicable for security purposes \cite{bi2016leverage}. For example, TFET can make the cryptographic processors to be more resilient toward side-channel attack. However, it should not be neglected that these properties can come to the aid of an attacker as well. An adversary may find a crack to cause performance degradation, functionality failure, acceleration of reliability issues and so forth. 

Therefore, development of novel VLSI testing and security checking techniques is mandatory with focus on these emerging devices. This work proposes a built-in-self-test architecture for security checking of the magnetic tunnel junction (MTJ) device under malicious process variations attack, in Section 2. Also, a general identification technique is presented to detect any abnormality in the behavior and activities of the employed circuitries within the MTJ testing architecture, in Section 3. We conclude the paper in Section 4.

\section{Testing and Security Checking of Magnetic Tunnel Junction}

\begin{figure}[htbp]
	\centering
	\includegraphics[width=3.5in]{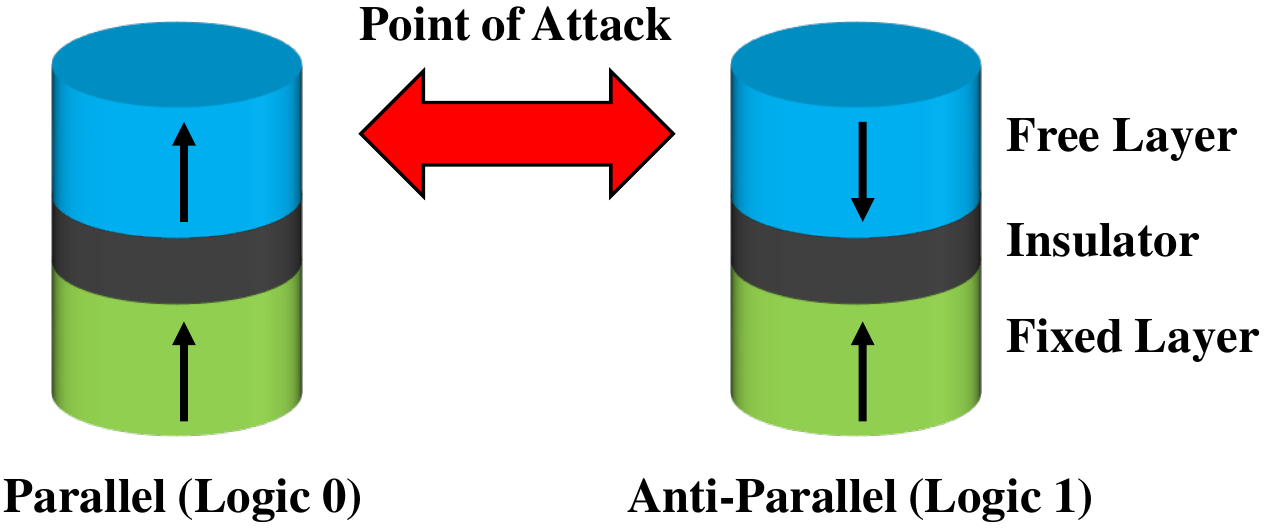}
	\caption{The perpendicular magnetic anisotropy-based magnetic tunnel junction device structure.}
	\label{fig:fig_1}
\end{figure}

\begin{figure}[htbp]
	\centering
	\includegraphics[width=3.5in]{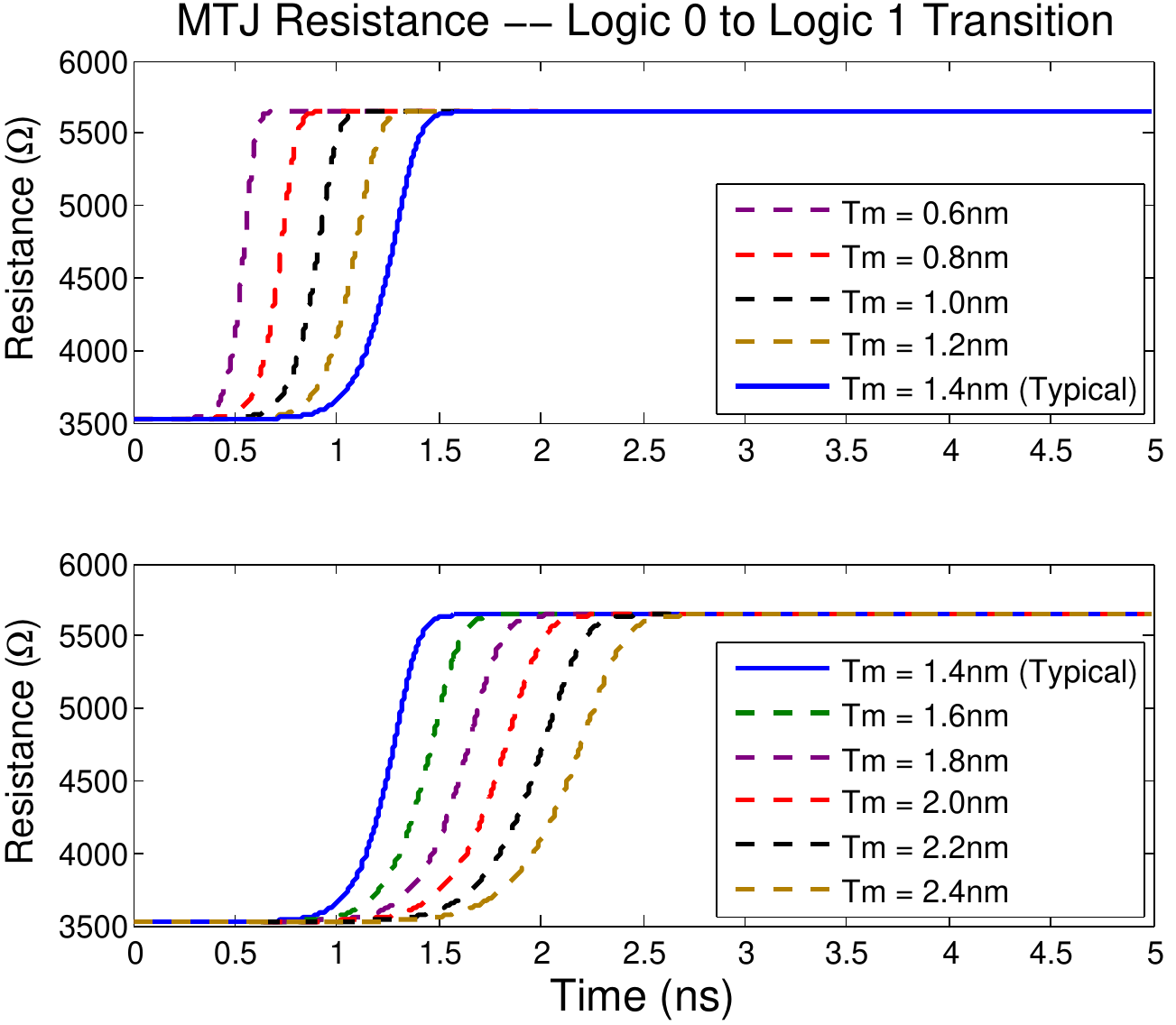}
	\caption{The free layer thickness malicious variations impact on the MTJ device resistance.}
	\label{fig:fig_2}
\end{figure}

Spintronics is the foundation of the next generation of memory technologies with having superior features such as energy efficiency, speed, and density in compare to the traditional memory technologies. Magnetic tunnel junction (MTJ) is the basic storage device in the Spintronics field that provides data non-volatility, fast data access, and low voltage circuit operation. These properties make this device a fit candidate for memory elements of the IoT devices \cite{ikeda2007magnetic}. However, the MTJ device may suffer from reliability issues \cite{wang2016compact} that can come to the aid of an attacker to perform malicious purposes. In this work, the impact of free layer thickness ($T_{m}$) malicious variations on the perpendicular magnetic anisotropy (PMA)-based MTJ device (shown in in Figure~\ref{fig:fig_1}) operation is analyzed using the SPICE models for magnetic tunnel junctions based on mono-domain approximation \cite{fong2013spice}. This attack can cause logical transitions of the MTJ device earlier or later than the expected time (displayed in Figure~\ref{fig:fig_2}), leading to an incorrect logical state sensing and its propagation throughout the system (especially in high clock frequencies).

One solution for preventing any possible timing failure caused by the reliability-related security issues is using run-time timing errors detection and possibly correction in order to keep a processing core performance close to its golden performance. According to this solution, a built-in-self-test module for reliability and security (BIST-RS) analysis is included inside the original design. The BIST-RS functionality can be classified to: (a) error detection; (b) error prediction; and (c) error masking. The BIST-RS functionality in "error detection" is described as monitoring the signals of logical paths for transitions after the clock edge and flagging a possible error.

\begin{figure}[htbp]
	\centering
	\includegraphics[width=3.5in]{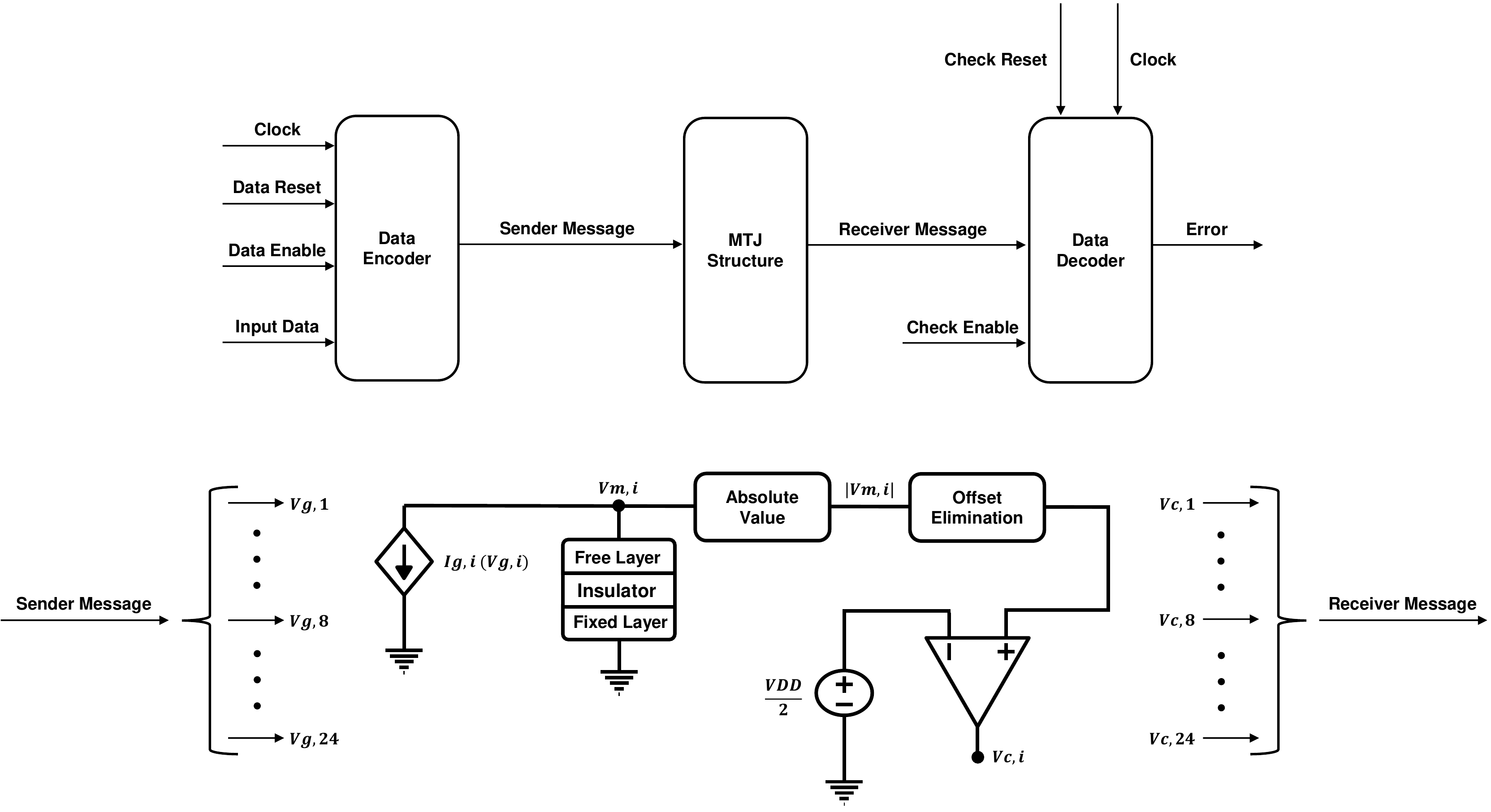}
	\caption{The BIST-RS architecture for reliability and security analysis of the magnetic tunnel junction device.}
	\label{fig:fig_5}
\end{figure}

In here, a BIST-RS architecture for analyzing the reliability and security of the MTJ device under malicious process variations attack is presented that is shown in Figure~\ref{fig:fig_5}. This architecture consists of three main elements: Data Encoder, MTJ Structure (i.e. an array of the MTJ cells), and Data Decoder. The data encoder has the responsibility of capturing the applied test pattern, calculating its fingerprint, and constructing the sender message. The MTJ structure is a physical transmission medium (i.e. the communication channel) with the functionality of correctly conveying data to the receiver. A healthy MTJ structure doesn't change the information and provides them to the data decoder on time. A single malicious MTJ cell (i.e. when the value of its free layer thickness is outside the acceptable range of variations) can change the conveying information. Regarding the logical state of each MTJ device, it is stayed the same or a transition happens depending on its corresponding bit in the applied test pattern. The data decoder checks the receiver message and declares its integrity status using the error signal. If the logical state of the error signal is high, it implies that the MTJ structure is not reliable/healthy and vice versa.

\begin{figure}[!h]
	\centering
	\includegraphics[width=3.5in]{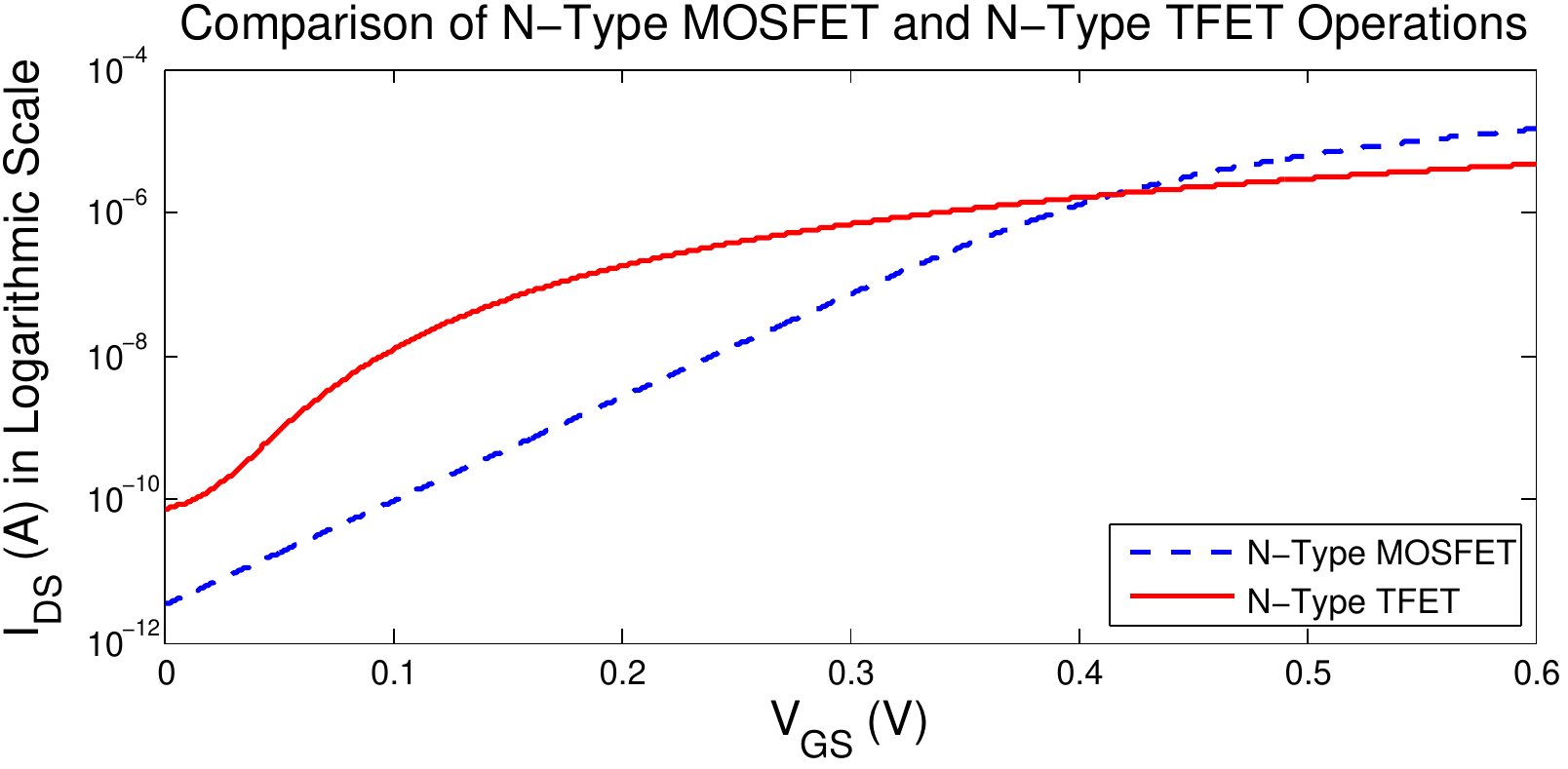}
	\caption{The comparison between the drain-source current versus gate-source voltage curves of the n-type MOSFET and the n-type TFET.}
	\label{fig:fig_16}
\end{figure}

The data encoder and the data decoder are demonstrated practically through hardware implementation of the cyclic redundancy check (CRC) using TFET technology. We use 20 nm AlGaSb/InAs tunnel field effect transistor (TFET) technology (provided in the Universal TFET model 1.6.8 \cite{nanoHUB_org31}) for implementation. TFET provides steeper sub-threshold slope (i.e. smaller than 60 mV/dec) \cite{fan2016challenges} and is described as a gated p-i-n (i.e. the hole-dominant region, the intrinsic (pure) region, and the electron-dominant region) diode that has asymmetrical doping structure and operates under reverse-bias condition. The steeper sub-threshold slope of the TFET device helps to further down scale the supply voltage and reduce the leakage currents substantially, which makes it an excellent candidate to achieve low energy consumption for the IoT applications. The comparison between the drain-source current ($I_{DS}$) versus gate-source voltage ($V_{GS}$) curves of the n-type MOSFET and the n-type TFET is shown in Figure~\ref{fig:fig_16}. For simulating this plot, both devices have the same width and length of 20 nm and are connected to the supply voltage of 0.6 V. As it can be seen from the figure, the TFET device turns ON and goes to its saturation region at a smaller value of the gate-source voltage in compare to the MOSFET device. Thus, the TFET technology is favorable for low voltage design.

\begin{figure}[!h]
	\centering
	\includegraphics[width=3.5in]{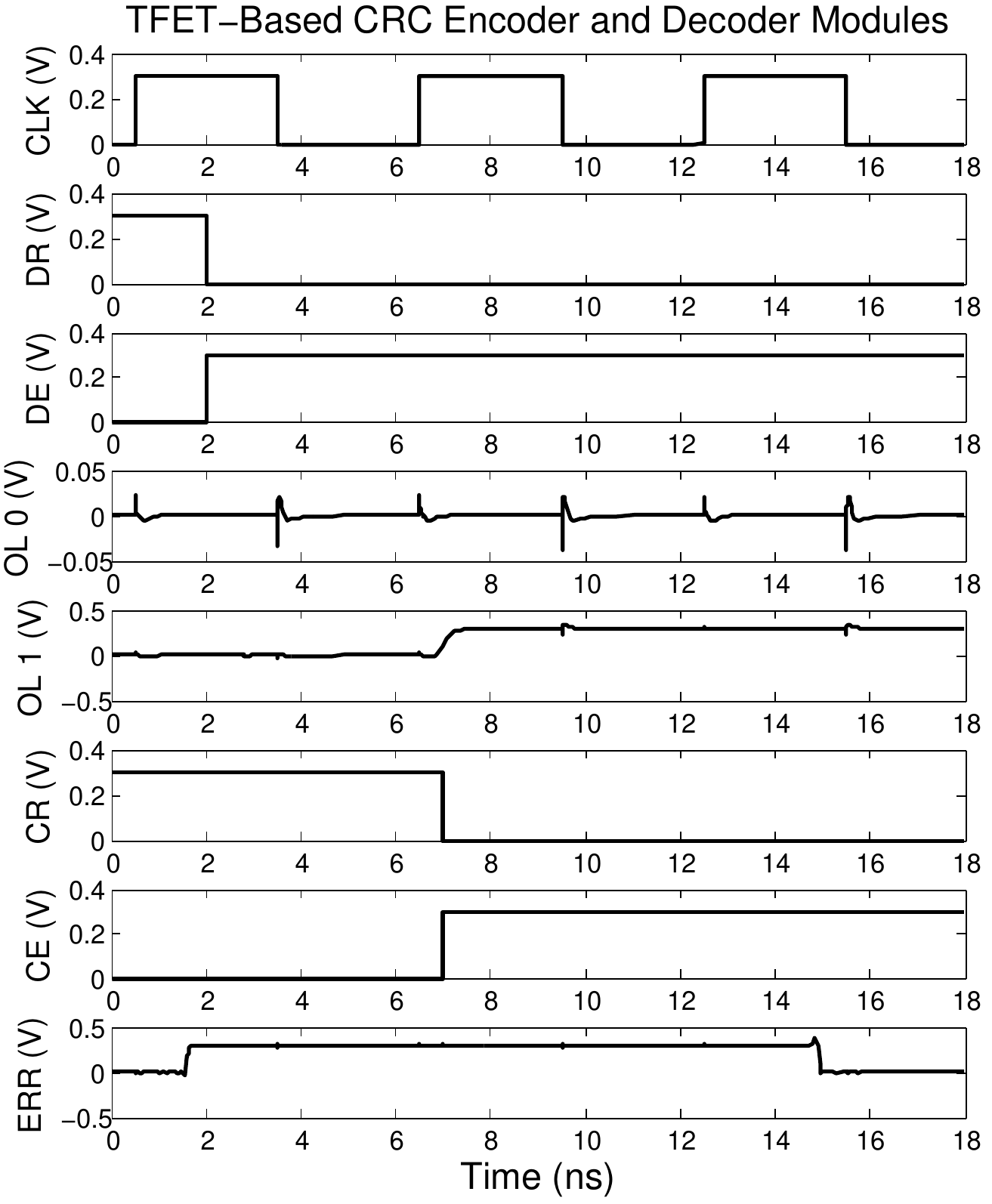}
	\caption{The TFET-based data encoder and data decoder input and output signals.}
	\label{fig:fig_6}
\end{figure}

Cyclic redundancy check is an error detection code that is used for authentic data transmission between a source and a destination \cite{CRC_2016}. The input and output signals of the data encoder and the data decoder can be seen in Figure~\ref{fig:fig_6}. The actual names of Y-axis labels are Clock, Data Reset, Data Enable, Output Logic 0, Output Logic 1, Check Reset, Check Enable, and Error in order. The clock signal has the period of 3 ns and the width of 6 ns. The reset mechanism of the encoder can be active before the arrival of the second clock cycle positive edge. Once it is disabled, the test pattern is applied and the data capturing signal is enabled. The constructed message is provided in the middle of the second clock cycle and around 7.5 ns. The fourth and fifth plots in Figure~\ref{fig:fig_6} show the example logic zero and logic one of the encoder output signals. Resetting the decoder element can be continued until the arrival of the third clock cycle positive edge. During this period, all of the flip-flops of the receiver remainder register are set to logic one. After the arrival of the third clock cycle positive edge, the error signal goes to logic zero or stays at logic one depending on the delivered message integrity status.

\begin{figure}[!h]
	\centering
	\includegraphics[width=3.5in]{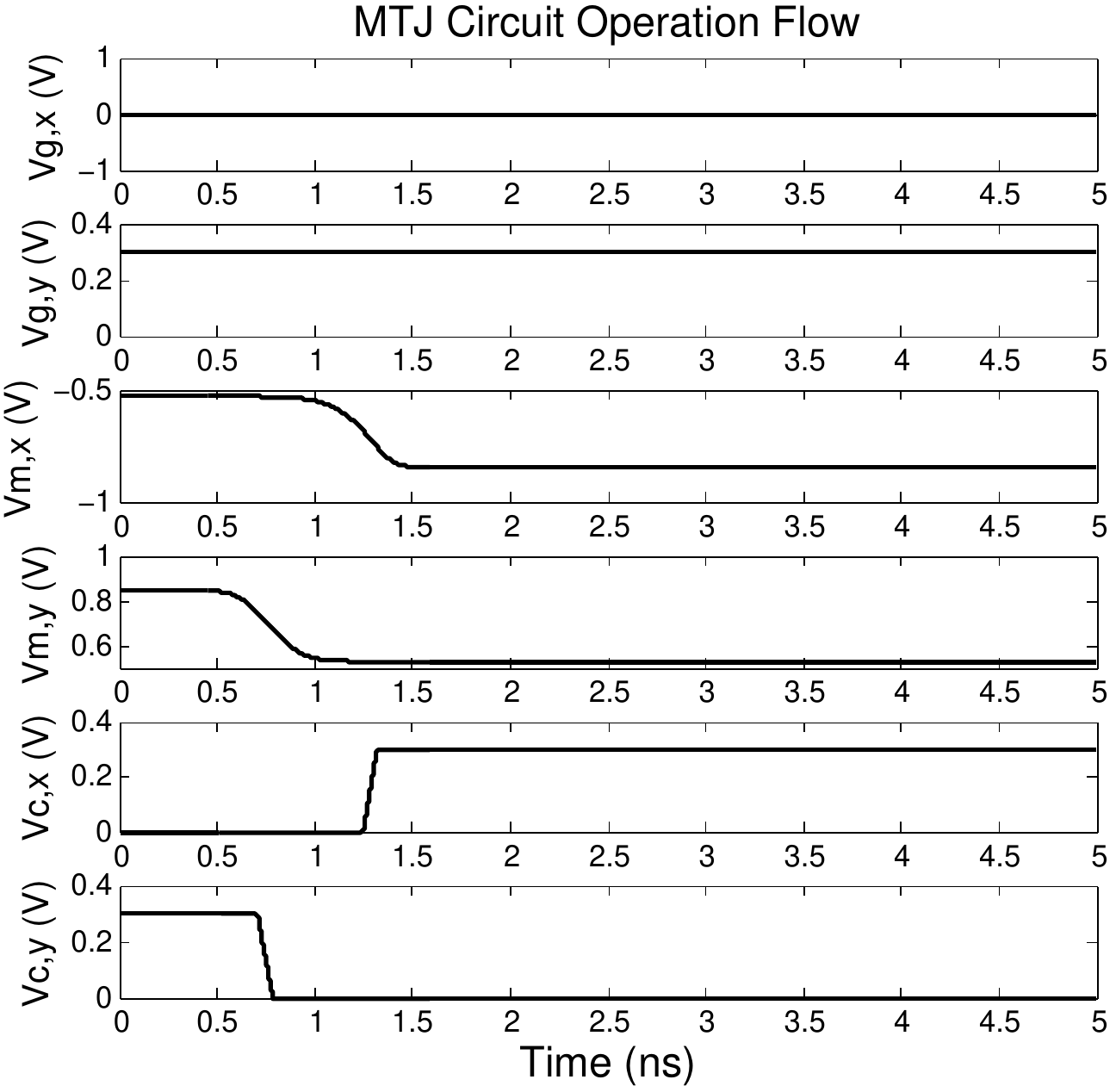}
	\caption{The operation flow of the MTJ driving and sensing circuit.}
	\label{fig:fig_7}
\end{figure}

Each cell in the MTJ structure contains a driving and sensing circuit for its magnetic tunnel junction,  which is shown in the bottom of Figure~\ref{fig:fig_5}. The operation of this circuit can be described in this way: (a) converting the voltage signal of a bit in the message to the current signal; (b) applying the current signal to the magnetic tunnel junction under test; and (c) finding the absolute value of the voltage signal at the free layer terminal since the voltage polarization is different between the MTJ logic transitions; (d) eliminating the signal offset to make sure that it is symmetric; and (e) comparing it to half of the supply voltage to construct the output signal based on the corresponding voltages of the logical states. Figure~\ref{fig:fig_7} indicates the circuit operation flow for zero-to-one and one-to-zero logic transitions using the inputs $V_{g,x}$ and $V_{g,y}$ respectively.

\begin{figure}[!h]
	\centering
	\includegraphics[width=3.5in]{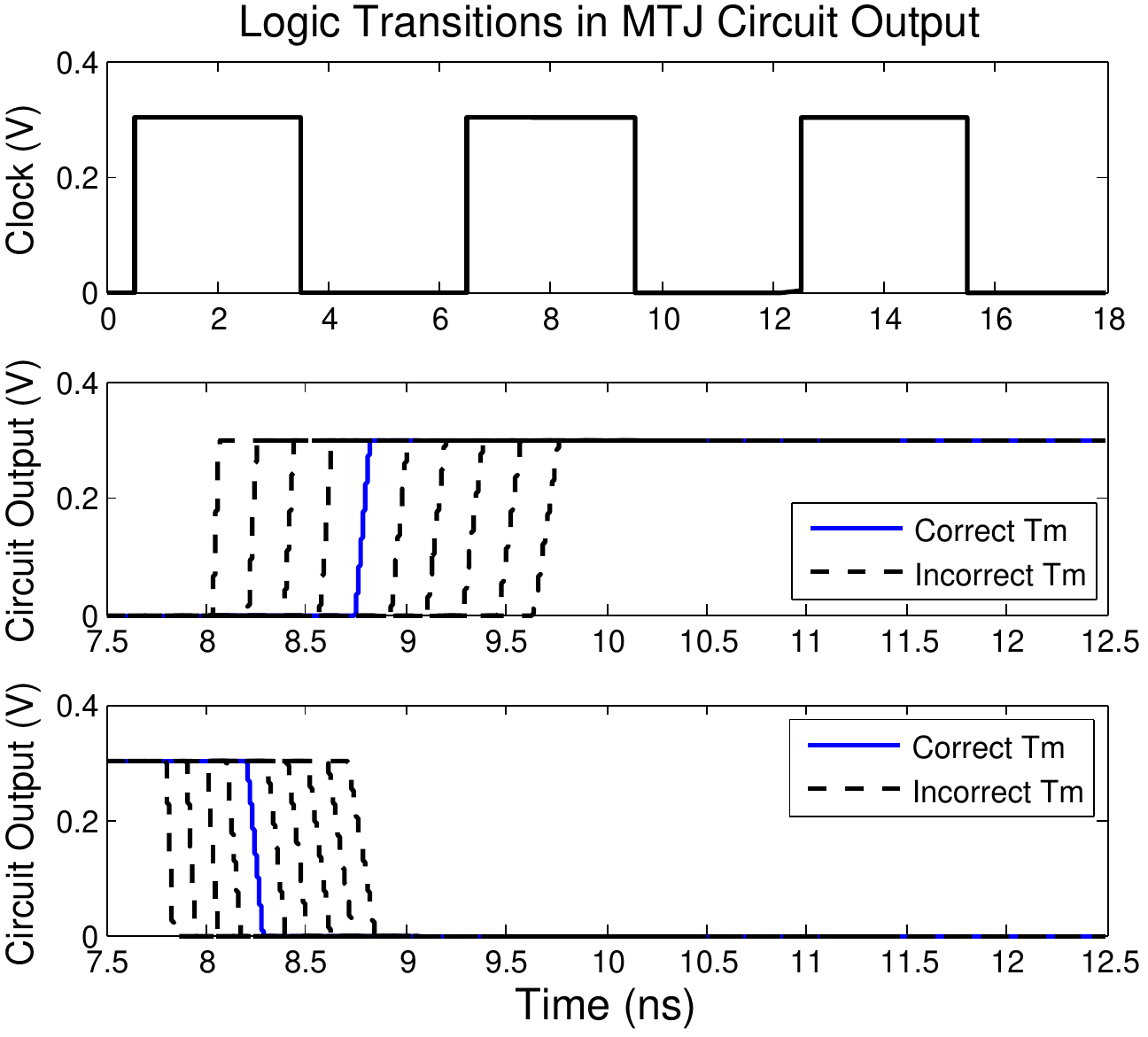}
	\caption{The MTJ driving and sensing circuit output under malicious free layer thickness variations.}
	\label{fig:fig_8}
\end{figure}

Now, the defense mechanism of the proposed BIST-RS architecture in confronting the malicious free layer thickness ($T_{m}$) variations is discussed. As the first and most important step, the BIST-RS clock frequency should be set to the desired clock frequency for the main circuit exactly. Then, the BIST-RS is turned ON and different test patterns are applied to the data encoder. Next, the data decoder captures the message at the clock cycle positive edge and evaluates its integrity. The message content might be wrong due to possible transition delay fault(s) caused by the infected MTJ(s). An infected MTJ is found by notification from the data decoder error signal. The illustration of this concept is shown in Figure~\ref{fig:fig_8}. As it can be observed, all the considered free layer thickness possibilities for zero-to-one and one-to-zero logic transitions are completed in the duration of 7.5 ns to 9.76 ns and 7.5 ns to 8.85 ns respectively, which are before the arrival of the third clock cycle positive edge. So, the malicious variations go undetected in this case. However, if this clock frequency is used for the original circuit all the times, then the attack doesn't have any impact on the IC functionality as well as its total performance. In reality, an IC might experience heavy workloads and high frequency computations during its lifetime. For those cases, the BIST-RS can be set to the clock frequency under test and the MTJ structure health is checked accordingly. The lack
of need for including additional memory resources for testing as well as detecting faults without necessity to propagate them throughout the circuit under test are the primary privileges of this architecture over the traditional testing and verification methods. Also, implementation of the encoder and decoder modules using the TFET technology brings less energy consumption and area occupation than its CMOS counterpart. The total power consumption and area of these modules are 0.5020 $\mu W$ and 1,930,400 $nm^2$.

\section{The BIST-RS for MTJ Under Attack}

Now, let's consider a scenario in which a malicious person is aware of the inserted BIST-RS inside the chip and aims to disrupt the testing and security checking process through manipulating the surrounding temperature or injecting a hardware Trojan inside the encryption/decryption module. For this scenario, we propose a general identification technique in this section, according to which any unusual behavior shown from the employed encryption/decryption module for MTJ testing can be discovered. Our technique performs the detection mechanism based upon the circuit analog signals rather than its digital data. Most of the detection techniques in the area of hardware security are developed in the digital domain, and the analog domain-based techniques have not been studied sufficiently \cite{bhunia2014hardware}.

In fact, the analog signals of an integrated circuit have unique variations, behavior, and features that can be used for detection, identification and monitoring purposes. The methodology of our technique can be divided into four steps: (a) choosing and applying a specific test pattern (i.e. based on its fault coverage capability) to the circuit under test and extracting an analog signal (i.e. the current signal in here), which is considered as the reference signal. This signal is correlated to the circuit properties. (b) automatic random selection of a certain number of test patterns (i.e. twenty in here), applying them to the circuit, and collecting all of the corresponding analog signals in order to build a dataset. Certain features may be extracted from the signals for the purpose of comparison in this step. (c) running a relational detector (i.e. the maximum of the absolute value of the cross correlation between the reference signal and a test signal) between the reference signal (i.e. obtaining when the circuit operates in normal condition) and all of the test signals inside the dataset in order to construct the "Evaluation Signal". (d) accepting or rejecting the evaluation signal depending on the detector threshold value (i.e. the mean of the reference evaluation signal) and its sensitivity, and calculating four basic statistical metrics for analyzing the detector performance. The four metrics for analysis of the detector performance are: True Positive (i.e. a signal is correctly rejected as not having originated from the original circuit), False Positive (i.e. a signal is wrongly rejected as not having originated from the original circuit), True Negative (i.e. a signal is correctly accepted as having originated from the original circuit), False Negative (i.e. a signal is wrongly accepted as having originated from the original circuit).

The presented approach is examined in two experiments. In the first experiment, four datasets are collected from the cyclic redundancy check data decoder circuit using the CMOS 20nm Predictive Technology Model (PTM) - Multi Gate (MG) technology \cite{PTM_2012}. The circuit operating conditions for these datasets are defined as: (1) normal condition; (2) process variations (i.e. changing the transistor length within $\pm$ 20\% range); (3) temperature variations (i.e. changing the temperature from 20$^{\circ}$C to 120$^{\circ}$C); and (4) malicious condition (i.e. a hardware Trojan is inserted inside the circuit). The designed hardware Trojan for this circuit is activated according to a logical AND function output with having the executed XOR function outputs on the CRC data decoder input pattern and the generated "Check Value" as its inputs, and its payload is the error signal malfunction. For the second experiment, only the normal condition and the malicious condition datasets are collected for the 32-bit KATAN block cipher \cite{de2009katan}, which its encryption and decryption modules can be used in the MTJ testing architecture as well. The inserted Trojan in the KATAN circuit has the duty of flipping the first and the last bits of the ciphertext, and is awakened according to a logical AND function output with having the executed XOR function outputs on a portion of the key and a portion of the plaintext as its inputs.

\begin{figure}[!h]
	\centering
	\includegraphics[width=3.5in]{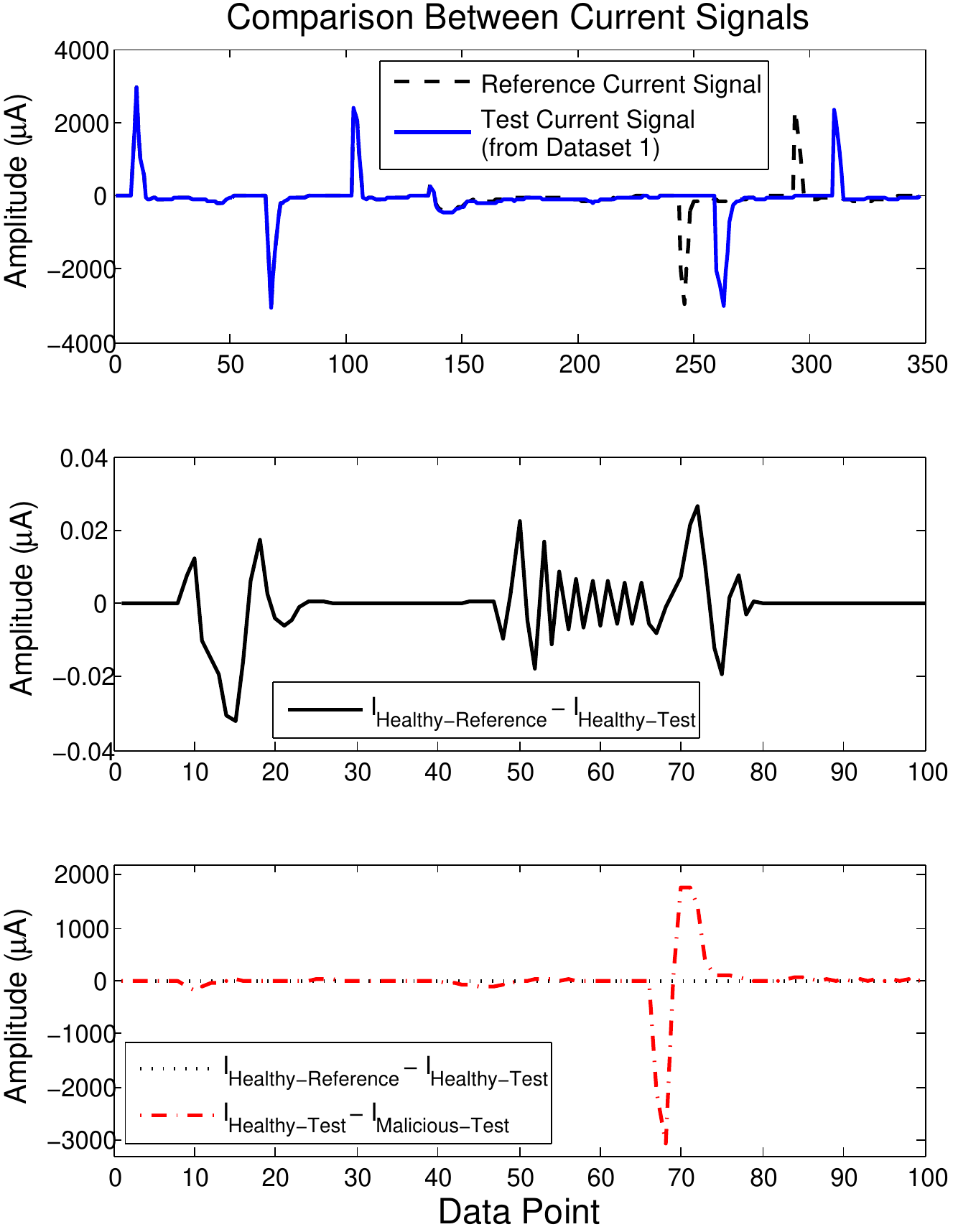}
	\caption{The comparison between the current signals of the healthy and the malicious CRC data decoders.}
	\label{fig:fig_11}
\end{figure}

\begin{figure}[!h]
	\centering
	\includegraphics[width=3.5in]{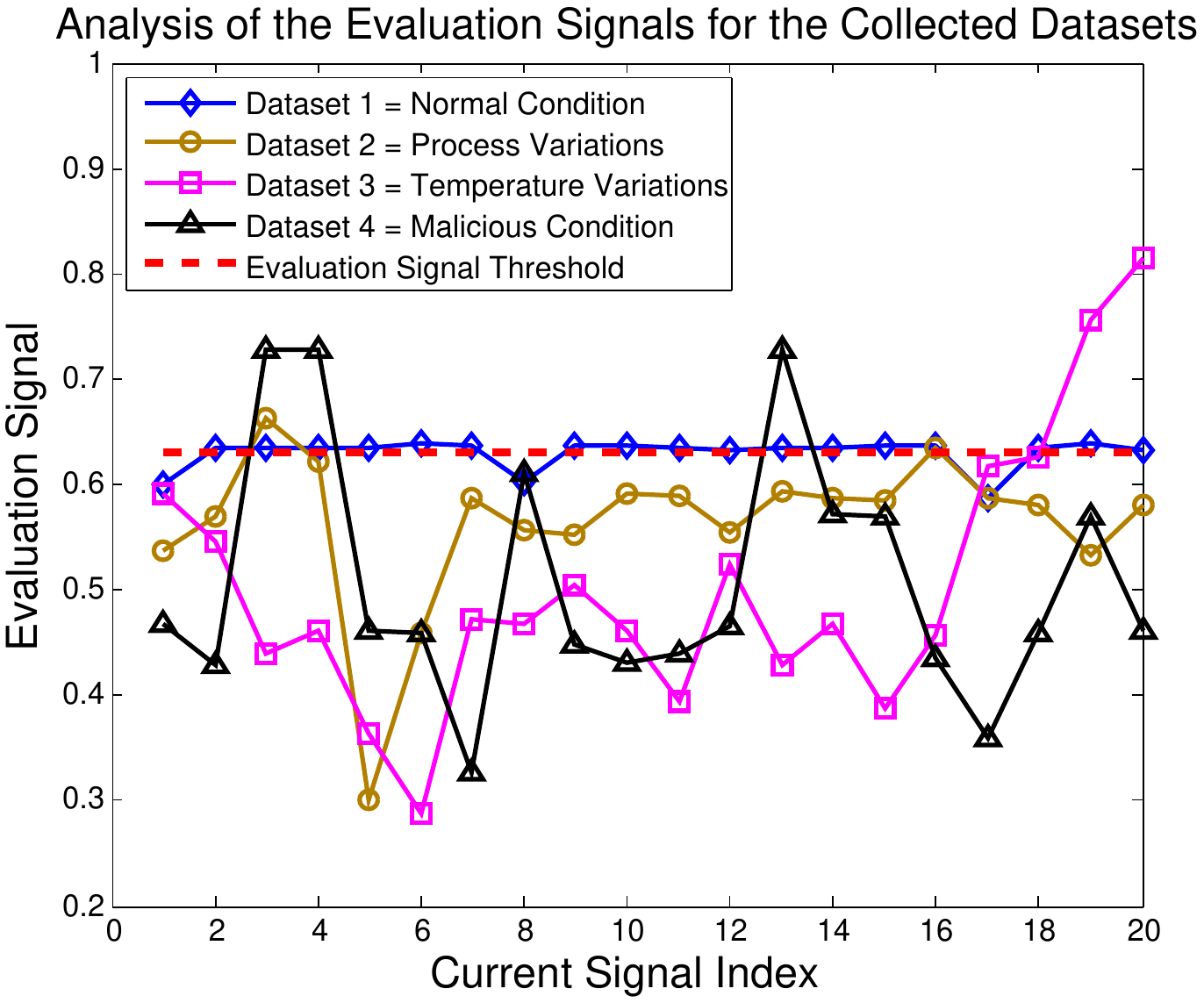}
	\caption{The analysis of the evaluation signals for the four datasets of the CRC data decoder.}
	\label{fig:fig_12}
\end{figure}

The top plot of Figure~\ref{fig:fig_11} shows the comparison between the current signals of the healthy CRC data decoder circuit (i.e. the normal condition) when the reference pattern and an arbitrary test pattern are applied. As it can be seen, the signals have the same trend and very well lie on each other at least for the first 100 data points. However, the middle plot of this figure demonstrates that there are still differences between the current signals for those data points. The comparison between the current signals of the healthy and the malicious CRC data decoders (i.e. the normal and malicious conditions respectively) when the same input pattern is applied can be observed in the bottom plot of the figure. Similarly, the signals have the same trend with minor differences, except some data points that the differences can be up to 40,000 times higher that is due to the hardware Trojan effect. Therefore, it can be interpreted that: (a) the circuit current signal is time and test pattern variant; and (b) the extracted current signals from applying two different input patterns have dissimilar variations at any given time, even if they have the same overall trend. In fact, these variations can cause a specific change in the level of the evaluation signal. The calculated evaluation signals for the four datasets of the CRC data decoder along with their threshold value are demonstrated in Figure~\ref{fig:fig_12}.

It can be comprehended from the figure that the evaluation signal of the first dataset is nearly stable and has the least amount of variations, which is due to not having remarkable variations among different applied input patterns in the circuit normal condition. The evaluation signals of the second and the third datasets have larger variations with having relatively constant behavior. The evaluation signal of the fourth dataset has the largest amount of variations and its behavior may be considered as abnormal in comparison with the other datasets. In the next step, the four basic statistical metrics for analysis of the detector performance with different levels of sensitivity are calculated. The results for these metrics using the four datasets of the CRC data decoder are presented in Table~\ref{tab:tab_2}.

\begin{table}[htbp]	
	\centering
	\includegraphics[width=3.5in]{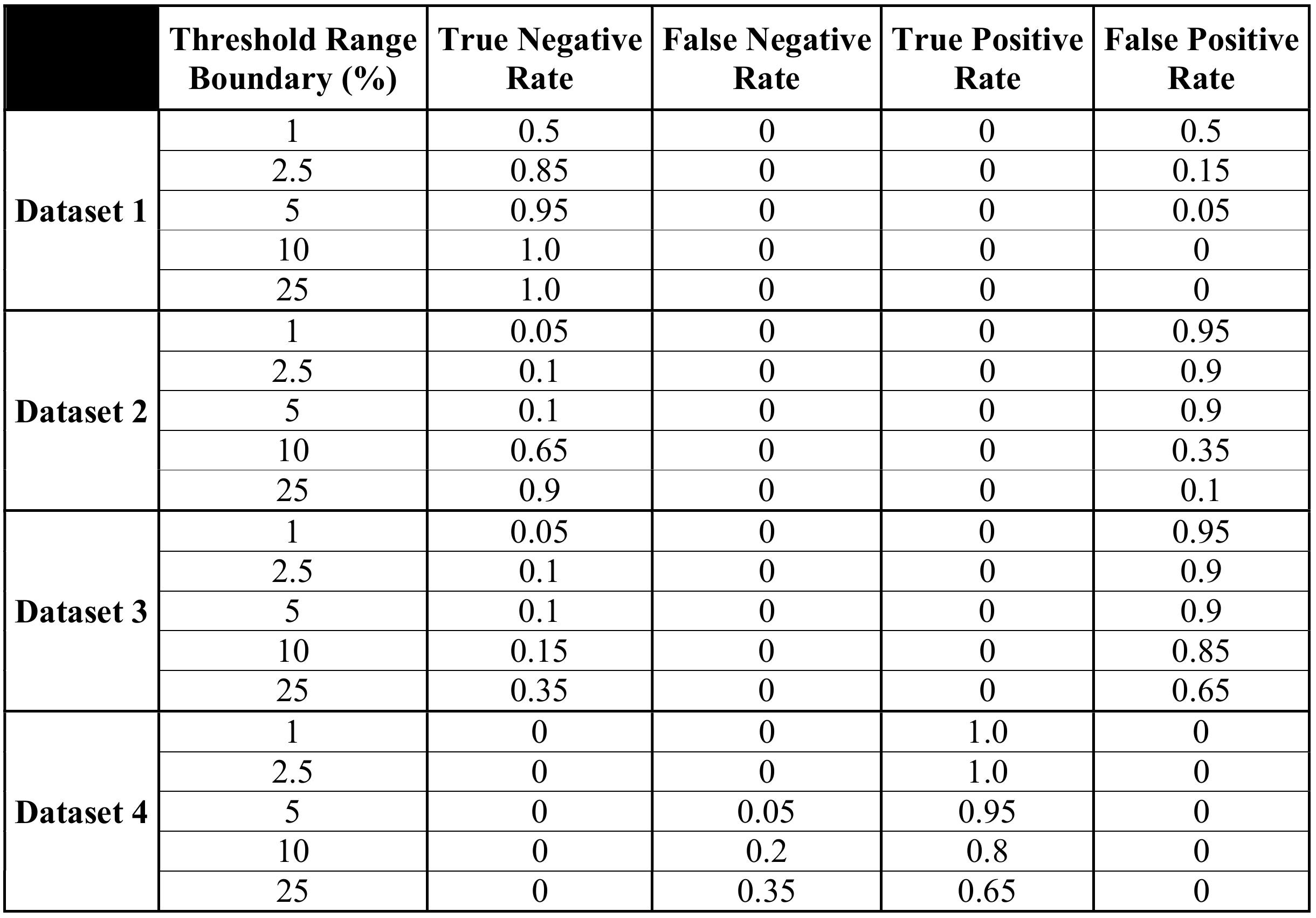}
	\caption{The analysis of the detector performance with different levels of sensitivity for the CRC data decoder.}
	\label{tab:tab_2}
\end{table}

According to the definitions of the four basic statistical metrics, the detector shows perfect performance in identification of the circuit in the normal condition as well as detection of the hardware Trojan. Also, it demonstrates a good performance in identifying the circuit when the variations of the process technology and the temperature are acceptable. Similar performance capability can be observed from the detector in identification of the KATAN block cipher circuit, which is shown in Table~\ref{tab:tab_3}.

\begin{table}[htbp]	
	\centering
	\includegraphics[width=3.5in]{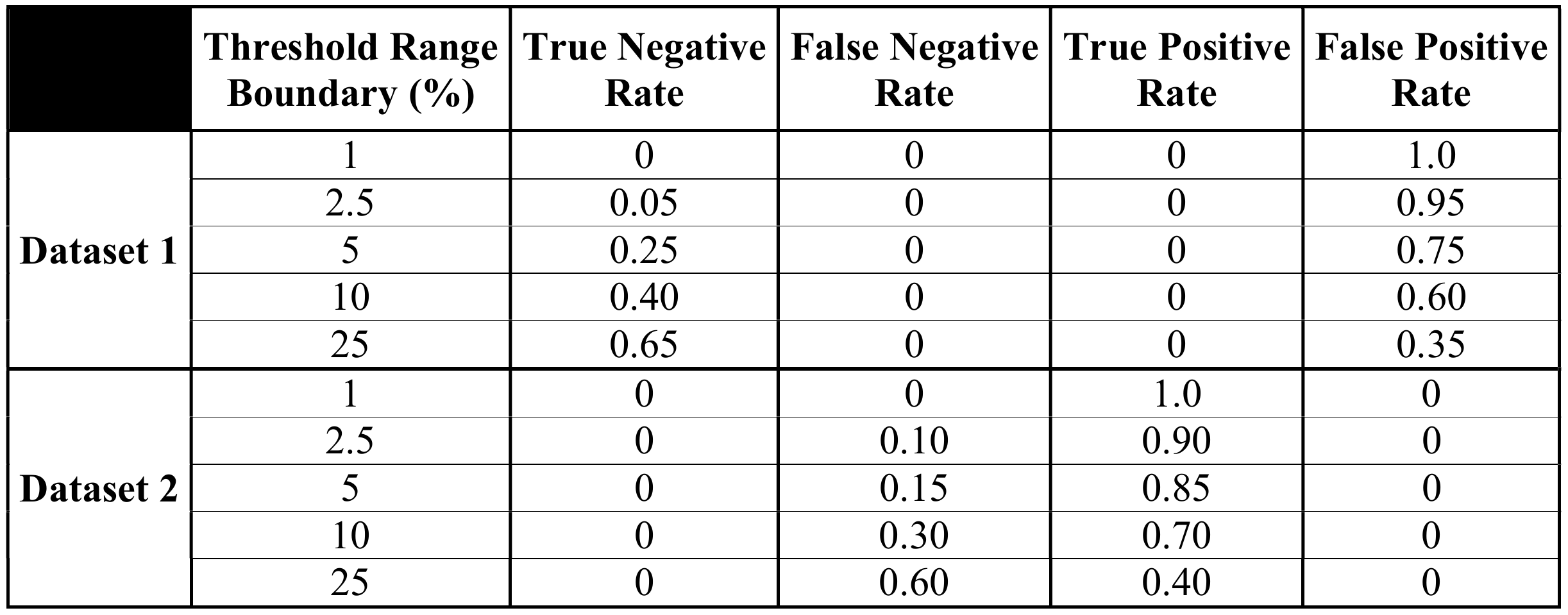}
	\caption{The analysis of the detector performance with different levels of sensitivity for the KATAN block cipher.}
	\label{tab:tab_3}
\end{table}

\section{Conclusion}

In this paper, we propose a built-in-self-test architecture for security checking of the MTJ device under malicious process variations attack. The architecture consists of three main elements: sender, physical transmission medium (i.e. an array of the MTJ cells), and receiver. A healthy array of MTJ cells doesn't change the sent information and delivers them to the receiver on time for integrity checking. The lack of need for including additional memory resources for testing as well as detecting faults without necessity to propagate them throughout the circuit under test are the primary privileges of this architecture over the traditional testing and verification methods. Also, a general identification technique is presented to discover any abnormal behavior and activity shown from the employed circuitries within the architecture. According to this technique, the existing features inside the current signal of a circuit under test can be used in order to identify it in different conditions, distinguish it from different circuits, and detect its possible infection caused by a hardware Trojan. The experimental results show that the technique has adequate performance in identifying the circuit under test in normal and malicious conditions as well as under typical process and temperature variations.





\bibliographystyle{IEEEtran}
\bibliography{IEEEabrv,./Reference_Paper}

\begin{IEEEbiographynophoto}{Shayan Taheri}
	received the B.S. degree in Electrical Engineering from the Shahid Beheshti University (National University of Iran), Tehran, Iran, and the M.S. degree in Computer Engineering from the Utah State University, Logan, UT, USA, in 2013 and 2015, respectively. He is currently pursuing the Ph.D. degree in Electrical Engineering at the University of Central Florida, Orlando, FL, USA. His research interests and experiences include the applications of new transistor and memory technologies in secure and low power VLSI design, hardware Trojan design and analysis for the Internet of Things (IoT) devices, leveraging signal processing in hardware security, and VLSI Testing and Verification.
\end{IEEEbiographynophoto}

\begin{IEEEbiographynophoto}{Jiann-Shiun Yuan}
	received the M.S. and Ph.D. degrees from the University of Florida, Gainesville, in 1984 and 1988, respectively. In 1988 and 1989 he was with Texas Instruments Incorporated, Dallas, for CMOS DRAM design. Since 1990 he has been with the faculty of the University of Central Florida (UCF), Orlando, where he is currently a Professor and Director of NSF Multi-functional Integrated System Technology (MIST) Center. He is the author of three textbooks and 300 papers in journals and conference proceedings. He supervised twenty-three Ph.D. dissertations, thirty-two M.S. theses, and five Honors in the Major theses at UCF. Since 1990, he has been conducting many research projects funded by the National Science Foundation, Intersil, Jabil, Honeywell, Northrop Grumman, Motorola, Harris, Lucent Technologies, National Semiconductor, and state of Florida.
	
	Dr. Yuan is a member of Eta Kappa Nu and Tau Beta Pi. He is a founding Editor of the IEEE Transactions on Device and Materials Reliability and a Distinguished Lecturer for the IEEE Electron Devices Society. He was the recipient of the 1995, 2004, 2010, and 2015 Teaching Award, UCF; the 2003 Research Award, UCF; the 2003 Outstanding Engineering Award, IEEE Orlando Section, the Excellence in Research Award at the full Professor level of the College of Engineering and Computer Science in 2015, and the Pegasus Professor Award, highest academic honor of excellence at UCF, in 2016.
\end{IEEEbiographynophoto}

\end{document}